\documentclass[twocolumn,preprintnumbers,superscriptaddress,amsmath,amssymb,pra]{revtex4-2}
\usepackage{amsmath,amssymb,graphicx,epstopdf,bm,upgreek, natbib}
\usepackage{graphicx}
\usepackage{color}
\definecolor{OliveGreen}{rgb}{0,0.6,0}
\usepackage{dcolumn}
\usepackage{bm}
\usepackage{amssymb,float}
\usepackage{cancel}
\usepackage{siunitx}
\usepackage{bbold}
\usepackage[colorlinks,bookmarks=false,citecolor=blue,linkcolor=blue,urlcolor=blue]{hyperref}
\usepackage{physics}
\usepackage{mathtools}
\usepackage{ulem}

\begin{document}
\title{Topological disclination states and charge fractionalization \\
  in a non-Hermitian lattice}

\author{Rimi Banerjee}
\affiliation{School of Physical and Mathematical Sciences, Nanyang Technological University,
Singapore 637371, Singapore}

\author{Subhaskar Mandal}
\affiliation{School of Physical and Mathematical Sciences, Nanyang Technological University,
Singapore 637371, Singapore}

\author{Yun Yong Terh}
\affiliation{School of Physical and Mathematical Sciences, Nanyang Technological University,
Singapore 637371, Singapore}

\author{Shuxin Lin}
\affiliation{School of Physical and Mathematical Sciences, Nanyang Technological University,
Singapore 637371, Singapore}

\author{Gui-Geng Liu}
\affiliation{School of Physical and Mathematical Sciences, Nanyang Technological University,
Singapore 637371, Singapore}

  \author{Baile Zhang}
\email{blzhang@ntu.edu.sg}
\affiliation{School of Physical and Mathematical Sciences, Nanyang Technological University, Singapore 637371, Singapore}
\affiliation{Centre for Disruptive Photonic Technologies, Nanyang Technological University, Singapore, 637371, Singapore}

 \author{Y. D. Chong}
\email{yidong@ntu.edu.sg}
\affiliation{School of Physical and Mathematical Sciences, Nanyang Technological University, Singapore 637371, Singapore}
\affiliation{Centre for Disruptive Photonic Technologies, Nanyang Technological University, Singapore, 637371, Singapore}


\begin{abstract}    
  We show that a non-Hermitian lattice with a disclination can host topological disclination states that are induced by on-site gain and loss.  The disclination states are inherently non-Hermitian as they do not exist in the limit of zero gain/loss.  They arise from charge fractionalization in the non-Hermitian lattice, which we establish using non-Hermitian Wilson loops calculated with biorthogonal products.  The model can be realized using an array of optical resonators, with the emergence of the topological disclination states manifesting as an abrupt shift in emission intensity and frequency upon tuning the gain/loss level.
\end{abstract} 

\maketitle


{\textit{Introduction}---}In discrete lattices, dislocations and disclinations are elementary defects that cannot be removed by local deformations, due to their global topological structure \cite{Mermin1979, Lusk2008Nano, Kim2009Large, Wei2012The, Wu2013Density, Kosterlitz2017Nobel}.  They are known to play an important role in many condensed-matter phenomena, such as the melting of two-dimensional solids \cite{Kosterlitz2017Nobel}.  In topological materials \cite{Hasan2010Colloquium}, such defects have a special significance: they allow us to formulate ``bulk-defect correspondences'', generalizing the bulk-boundary correspondences that usually serve as physical signatures for bulk band topology \cite{Ran2009One, Teo2013Existence, Ruegg2013Bound, Benalcazar2014Classification, Li2018Topological, Teo2010Kane, TianheFractional2020, Liu2021Bulk, Peterson2021Trap, Wang2020Observation, TianheFractional2020, Wang2021Vortex, Xue2021Observation, Max2021Bulk, Deng2022Observation, Xie2022Photonic, Lin2023Topo, Hwang2023Vortex}.  In certain higher-order topological insulators (HOTIs) \cite{Benalcazar2017Quantized, Bradlyn2017Topological, PetersonA2020, Noh2018Topological, Ezawa2018Higher, Xie2018Second, Marc2018Observation, Xue2019Acoustic, Li2020Higher, Imhof2018Topolectrical}, bulk-defect correspondences may be used to probe topological properties that are hard to access through boundary measurements \cite{Liu2021Bulk,Peterson2021Trap}.  Aside from the appearance of localized defect states, some bulk-defect correspondences predict charge fractionalization, whereby a fraction of a unit charge polarization is localized at the defect.  This phenomenon, which results from an incompatibility between the charge quantization principle and lattice symmetries (a filling anomaly), underpins the topology of HOTIs \cite{Benalcazar2017Quantized, Bradlyn2017Topological, PetersonA2020}.  Charge fractionalization and localized states have recently been observed experimentally in 2D photonic lattices with disclinations \cite{Liu2021Bulk, Peterson2021Trap}.

Theoretical analyses of topological materials, including most earlier studies of bulk-defect correspondences, usually take Hermiticity as a starting assumption.  In recent years, however, there has been increasing interest in non-Hermitian (energy non-conserving \cite{Bender2007Reports, feng2017non, Wang2023review}) topological materials \cite{Lee2016Anomalous, Leykam2017Edge, shen2018topological, Gong2018Topological, Cerjan2019Experimental, ghatak2019new, torres2019perspective, Kawabata2019Symmetry, Parto2021Non, Bergholtz2021Exceptional}.  Not only do such systems pose the theoretically interesting challenge of formulating band topology without Hermiticity, but they are also of practical interest: in the classical-wave metamaterials commonly used to realize topological phases, loss and gain are often non-negligible \cite{Haldane2008Possible, Wang2009Observation, Lu2014Topological, Yang2015Topological, Nash2015Topological, Ningyuan2015Time, Albert2015Topological}.  For example, topological lasers, which are promising technological applications of topological states, are inherently non-Hermitian (NH) \cite{Jean2017Lasing, Parto2018Edge, Miguel2018Topological, zeng2020electrically, Dikopoltsev2021Topological, yang2022laser}.  Research into NH band topology has uncovered numerous surprises, such as NH topological phases distinct from any Hermitian counterpart \cite{Lee2016Anomalous, Leykam2017Edge, Weimann2017Topologically, Kawabata2018Anomalous, Bergholtz2021Exceptional}.  Recently, there have been studies showing that gain and loss can induce boundary states in 1D lattices \cite{Takata2018Pho, Liu2020Gain, Gao2020Observation} and corner states in 2D lattices \cite{Luo2019High, Gao2021Non-Hermitian}.  To our knowledge, however, topological bulk-defect correspondences have not yet been studied in the non-Hermitian regime.

In this paper, we demonstrate that a NH lattice containing a disclination can host topological disclination states associated with fractional (1/2) charge.  Unlike the previously-studied disclination states of Hermitian lattices \cite{Ran2009One, Teo2013Existence, Ruegg2013Bound, Benalcazar2014Classification, Li2018Topological, Wang2020Observation, TianheFractional2020, Wang2021Vortex, Xue2021Observation, Liu2021Bulk, Peterson2021Trap, Max2021Bulk, Deng2022Observation, Xie2022Photonic, Lin2023Topo, Hwang2023Vortex}, these NH disclination states are induced solely by on-site gain and loss, and are non-existent in the Hermitian limit.  The model consists of a 2D graphene-like honeycomb lattice with a disclination; the application of a specific pattern of gain/loss (imaginary mass terms) generates a bulk gap in the real part of the energy spectrum, in which the disclination states appear \cite{Xu2022Edge, Chen2021Comparative, Zhu2021Single, Li2022Effective}.  Previously, gain/loss-induced topological states have been found in NH models like the Takata-Notomi model, a 1D lattice hosting NH midgap boundary states \cite{Takata2018Pho}, as well as a NH 2D HOTI with gain/loss-induced corner states \cite{Luo2019High}.  NH topological disclination states, however, have not yet been identified.  (There have been some studies of how lattice defects affect the non-Hermitian skin effect, which is a separate issue \cite{SchindlerDislocation2021, Sun2021Geometric, Panigrahi2022Non-Hermitian, Bhargava2021Non}.)  This work also establishes NH charge fractionalization at a lattice defect, a phenomenon previously limited to Hermitian models \cite{TianheFractional2020, Liu2021Bulk, Peterson2021Trap}.  The charge fractionalization is established using Wilson loops derived from biorthogonal products, and also using a NH formulation of the density of states.  We will also discuss an alternative model that features NH disclination states unaccompanied by fractional charge.

\begin{figure*}
 \centering
\includegraphics[width=0.98\textwidth]{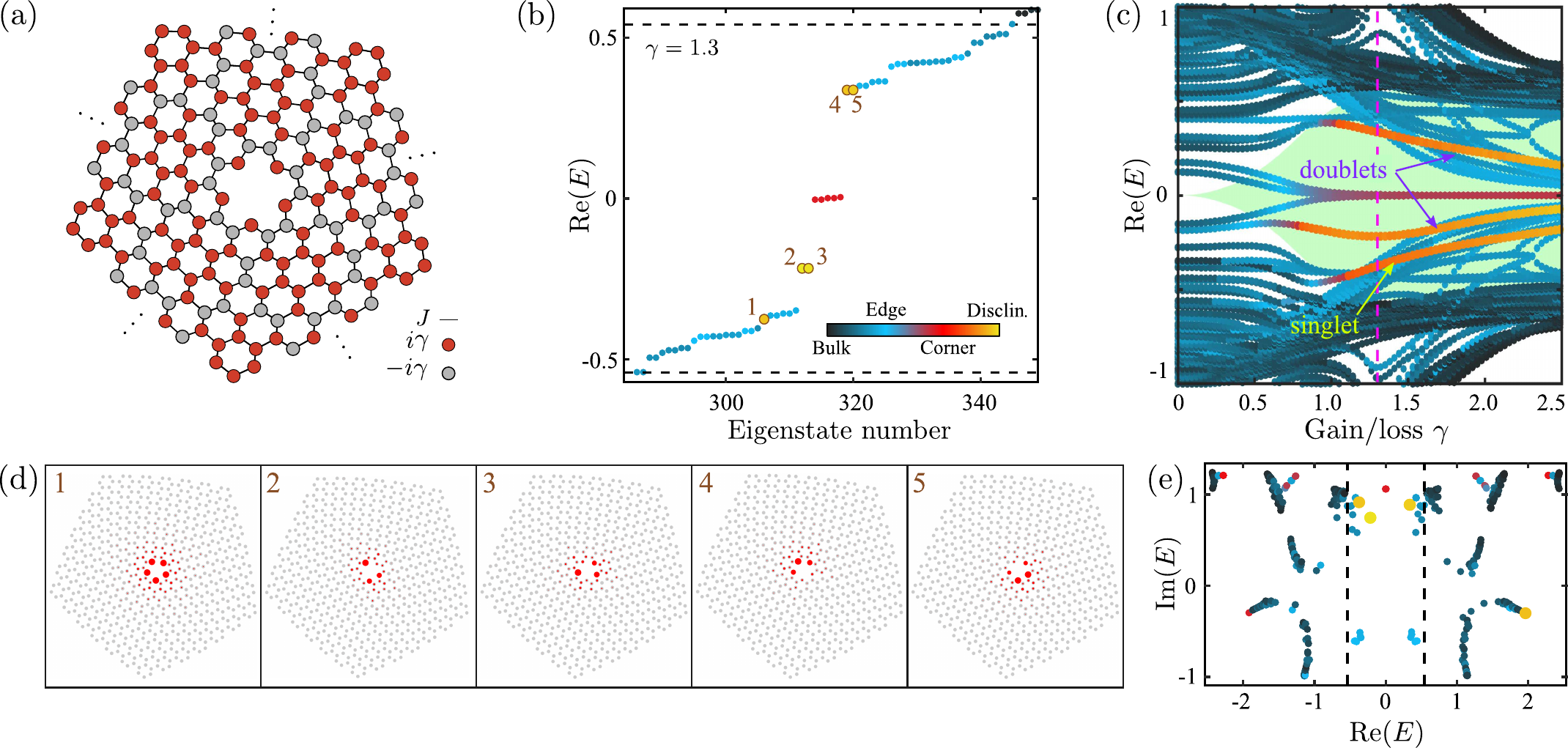}
\caption{(a) Schematic of a 2D NH lattice containing a disclination, with nearest-neighbor hopping $J$ (black links) and on-site gain $i\gamma$ (red circles) or loss $-i\gamma$ (gray circles). (b) $\mathrm{Re}(E)$ spectrum for the finite lattice in (a), but expanded to 630 sites.  Marker colors indicate the degree of localization to the bulk/edge/corners/disclination (see Supplemental Materials \cite{SuppMat}).  Horizontal dashes indicate the bulk gap obtained from Fig.~\ref{Fig:Main2}(a), in which lie the edge, corner, and disclination states. The energies labelled $1$--$5$ correspond to disclination states. (c) Plot of $\mathrm{Re}(E)$ versus $\gamma$, with the same marker color scheme as in (b).  The shaded green area corresponds to the bulk gap, and vertical magenta dashes indicate the reference gain/loss level $\gamma =1.3$. (d) Intensity distributions for the disclination states in (b).  (e) Complex energy spectrum for the disclination-bearing lattice of (b) and (d).  In all subplots, the hopping is $J=1$.}
\label{Fig:Main1}
\end{figure*}

\begin{figure}
\centering
\includegraphics[width=0.46\textwidth]{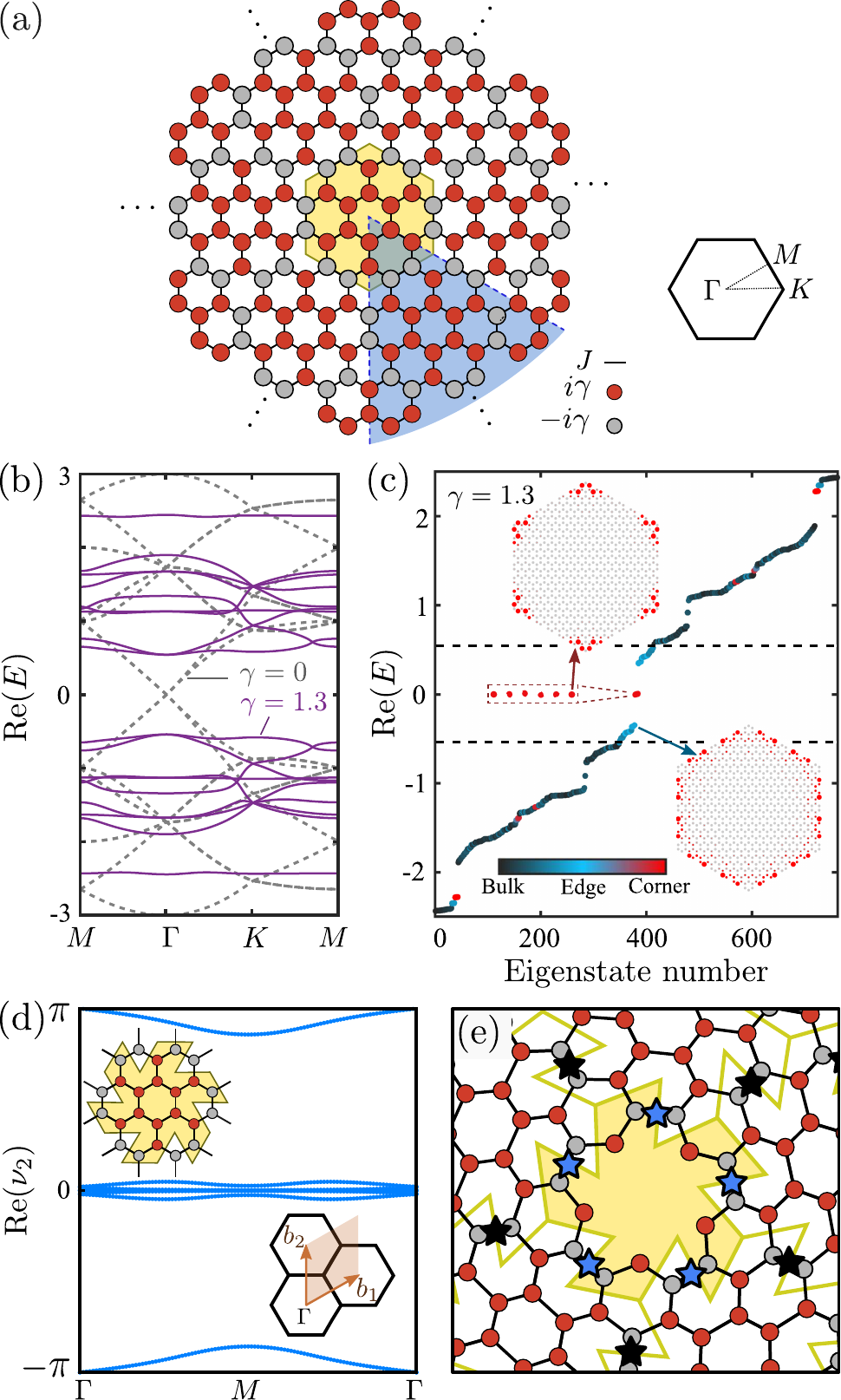}
\caption{(a) Schematic of a 2D NH honeycomb lattice with an 18-site unit cell (yellow region).  From this, Fig.~\ref{Fig:Main1}(a) can be generated by cutting out a $\pi/3$ sector (blue region).  (b) Real bulk band diagram for $\gamma = 0$ (gray dashes) and $\gamma = 1.3$ (purple lines). (c) $\mathrm{Re}(E)$ spectrum for a 762-site hexagonal sample with $\gamma = 1.3$.  Marker colors indicate the degree of localization to the bulk/edge/corners \cite{SuppMat}.  In the bulk gap (horizontal dashes), there are edge and corner states, with the latter pinned to $\mathrm{Re}(E) = 0$.  Inset: intensity ($|\psi|^2$) distributions for representative corner and edge states. (d) Wilson loops for the lowest nine bands at $\gamma=1.3$. Upper inset: the unit cell for this calculation.  Lower inset: layout of Brillouin zones and reciprocal lattice vectors $b_{1,2}$. (e) Positions of Wannier centers (black and blue stars) near the disclination.  Those adjacent to the disclination core (blue stars) yield a fractional charge of $1/2~(\mathrm{mod}~1)$.  In all subplots, $J=1$.}
\label{Fig:Main2}
\end{figure}

Finally, we study the prospects for realizing such a NH lattice using pumped optical resonators.  In such a setting, the gain/loss-induced topological disclination mode has a distinctive experimental signature in the form of an increase in emission intensity and abrupt shift in peak frequency as the gain/loss level is tuned.  The unique properties of NH topological disclination states, which are gain/loss-induced and yet possess a level of robustness due to their topological origins, may eventually be useful for designing novel lasers and related devices.

\textit{Model}---We begin with the 2D NH lattice shown in Fig.~\ref{Fig:Main1}(a), which has real nearest-neighbour hopping $J$ and bears a disclination at its core. Each site is assigned an imaginary mass corresponding to gain ($i\gamma$, red circles) or loss ($-i\gamma$, gray circles), where $\gamma \ge 0$ is the gain/loss level \cite{Bender2007Reports}.  The Hamiltonian is
\begin{equation}
  H = \sum_{n} i\gamma_n a^\dagger_n a_n + J \sum_{\langle nn'\rangle} \left(a^\dagger_{n'} a_n + \mathrm{h.c.}\right),
\end{equation}
where $\gamma_n = \pm \gamma$, $a_n^\dagger$ and $a_n$ are the creation and annihilation operators on site $n$, and $\langle nn' \rangle$ denotes nearest neighbors.  For $\gamma = 0$, this reduces to a Hermitian graphene lattice with a disclination \cite{Ruegg2013Bound}, which has no bulk gap and no edge, corner, or disclination states \cite{SuppMat}.

For nonzero $\gamma$, Fig.~\ref{Fig:Main1}(b) shows the real part of the spectrum.  The lattice is based on Fig.~\ref{Fig:Main1}(a) but expanded to 630 sites, with $\gamma = 1.3$ and $J = 1$.  There are several eigenenergies near the center of the gap, marked in blue and red, which turn out to be edge and corner states localized to the outer boundary of the sample \cite{SuppMat}.

We also see five eigenenergies, two doublets and a singlet, marked in yellow.  These turn out to be gain/loss-induced disclination states.  As shown in Fig.~\ref{Fig:Main1}(c), they emerge into the bulk gap at nonzero values of $\gamma$.  Their intensity distributions are plotted in  Fig.~\ref{Fig:Main1}(d), showing that they are localized to the vicinity of the disclination core.  The corresponding phase plots, presented in the Supplemental Materials \cite{SuppMat}, reveal the singlet state to have all five sites at the disclination core in-phase, whereas the doublets have more complicated phase distributions.  The distribution of eigenenergies in the complex $E$ plane is shown in Fig.~\ref{Fig:Main1}(e). Later, we show that these NH disclination states are also associated with a fractional charge of $1/2$. 

To understand these results, we consider the related disclination-free NH honeycomb lattice shown in Fig.~\ref{Fig:Main2}(a).  It has $C_6$ rotational  symmetry, with a unit cell (yellow hexagon) of 18 sites. From this lattice, we can generate the disclination lattice of Fig.~\ref{Fig:Main1}(a) through a Volterra (``cut-and-glue'') process, by removing a $\pi/3$ sector (blue-shaded region) and reconnecting the seams.  With this choice of gain/loss distribution, there is no seam or discontinuity after the Volterra process (there are other gain/loss distributions that are simpler and have smaller unit cells, but lack such a property).

Fig.~\ref{Fig:Main2}(b) shows the real part of the bulk band diagram for the NH lattice of Fig.~\ref{Fig:Main2}(a), with hopping $J = 1$.  In the Hermitian limit (i.e., $\gamma = 0$), the eigenenergies are real and gapless (gray dashes); this is simply the band diagram for graphene, with the original $K$ and $K'$ points folded onto $\Gamma$.  For $\gamma \ne 0$, a real line gap---i.e., a gap in the real part of the energy, $\mathrm{Re}(E)$ \cite{Bergholtz2021Exceptional}---opens up around $\mathrm{Re}(E) = 0$ (purple lines).  In Fig.~\ref{Fig:Main2}(c), we plot $\mathrm{Re}(E)$ for a finite hexagonal sample with 762 sites.  This spectrum exhibits the same real line gap as the bulk system, but with additional edge and corner states occupying the bulk gap.  The intensity distributions for two exemplary states are plotted in the inset of Fig.~\ref{Fig:Main2}(c).  Evidently, the lattice behaves much like a NH version of a HOTI \cite{Benalcazar2017Quantized}, with the nonzero gain and loss generating a real line gap as well as in-gap boundary states  \cite{Luo2019High, SuppMat}.

\textit{Topological characterization}---In Hermitian HOTIs, the topological states are tied to the deeper phenomenon of robust charge fractionalization \cite{TianheFractional2020, Liu2021Bulk, Peterson2021Trap}.  Here, we argue that the disclination in the present NH model induces a fractional charge of 1/2.

For the periodic lattice of Fig.~\ref{Fig:Main2}(a), we numerically calculate Wilson loops involving all nine bands situated below the real line gap.  We first adjust the shape of the unit cell to avoid the ambiguity of certain sites falling on the edges of the unit cell [Fig.~\ref{Fig:Main2}(d), upper inset].  Next, we discretize the first Brillouin zone along the reciprocal lattice vectors $b_1$ and $b_2$ [Fig.~\ref{Fig:Main2}(d), lower inset]. Starting from a given $k$, the Wilson loop matrices are \cite{Luo2019High,Gao2021Non-Hermitian}
\begin{align}
  W_{j} = F_j(k+N\Delta k_j) \cdots F_j(k+\Delta k_j)F_j(k),
\label{eq:EqW}
\end{align}
where $N$ is the number of discretization points in the $b_j$ direction, $\Delta k_j$ is the wavevector step, and $F_j(k)$ is a $9\times9$ matrix of biorthogonal products:
\begin{equation}
  F_j^{mn}(k) = \big\langle u_m^L(k+\Delta k_j) \big| u_n^R(k)\big\rangle.
\end{equation}
Here, $m,$ $n$ index bands below the line gap at $\mathrm{Re}(E) = 0$, and $u^{L/R}(k)$ denotes left/right eigenvectors of the Bloch Hamiltonian, which satisfy the biorthogonality relation $\bra{u_m^L(k)}\ket{u_n^R(k)}=\delta_{mn}$ \cite{Bender2007Reports}.

From the eigenvalues $\lambda_j$ of $W_j$, we retrieve the phases $\nu_j = -i \ln (\lambda_j)$.  Fig.~\ref{Fig:Main2}(d) plots $\mathrm{Re}(\nu_2)$ as the initial $k$ point varies along the $b_1$ direction.  Throughout this trajectory, $\mathrm{Im}(\nu_2)$ is close to zero. The plot for $\mathrm{Re}(\nu_1)$, as $k$ varies along $b_2$, behaves very similarly and is thus omitted.  The Wilson loop exhibits no overall winding, but crosses $\pm \pi$ an odd number of times.  This is the same behavior observed in the Hermitian Wu-Hu model, using standard rather than biorthogonal inner products to calculate $W_j$ matrices \cite{Wu2015Scheme, ProctorRobustness2020, Blanco2020Tutorial, Palmer2021Berry}.  That Hermitian model is known to belong to the 2D Stiefel-Whitney class and supports a HOTI phase with Wannier centers at the $3c$ Wyckoff position \cite{Palmer2021Berry, Ahn2019Stiefel, Xu2023Absence}.

Informed by these results, we take a close-up view of the disclination-bearing lattice from Fig.~\ref{Fig:Main1}(a).  As shown in Fig.~\ref{Fig:Main2}(e), the Wannier centers (black and blue stars) lie at the boundaries of the unit cells, contributing $1/2~(\mathrm{mod}~1)$ spectral charge to each adjacent cell \cite{TianheFractional2020, Liu2021Bulk}.  Since the boundary of the disclination core passes through an odd number of Wannier centers (blue stars), the disclination carries a fractional charge of $1/2~(\mathrm{mod}~1)$.  In Hermitian lattices, a similar argument for charge fractionalization at disclinations \cite{TianheFractional2020} has been verified in experiments on microwave metamaterials and photonic crystals \cite{Peterson2021Trap, Liu2021Bulk}.

Another way to identify disclination charge is to examine the local density of states \cite{Peterson2021Trap, Liu2021Bulk, PetersonA2020}.  For each site $j$, we define the expected density
\begin{equation}
  \langle \Pi_{jm} \rangle = \bra{\psi^L_m}\Pi_j \ket{\psi^{R}_m},
\end{equation}
where $\Pi_j=\ket{j}\bra{j}$ is a local projection operator and $|\psi^{L/R}_m\rangle$ are the biorthogonal $m$-th left/right eigenstates of the full lattice.  For our NH model, $\langle \Pi_{jm} \rangle$ is complex-valued \cite{Brody2014Biorthogonal, Kunst2018Biorthogonal, Bergholtz2021Exceptional}.  We define the disclination charge as
\begin{align}
  Q_{d}=\sum_{mj} \big| \langle \Pi_{jm} \rangle \big| \; (\text{mod~1}),
  \label{eq:EqQdis}
\end{align}
with $j$ summed over sites near the disclination core \cite{SuppMat}, and $m$ summed over all states below the bulk gap---i.e., $\mathrm{Re}(E)$ below the bottom dashes in Fig.~\ref{Fig:Main1}(b).  For the lattice of Fig.~\ref{Fig:Main2}(e), we obtain $Q_{d}=0.53$, close to the predicted value of $0.5$. This result is compatible with the previous Wannier center analysis.

\begin{figure}
\centering
\includegraphics[width=0.48\textwidth]{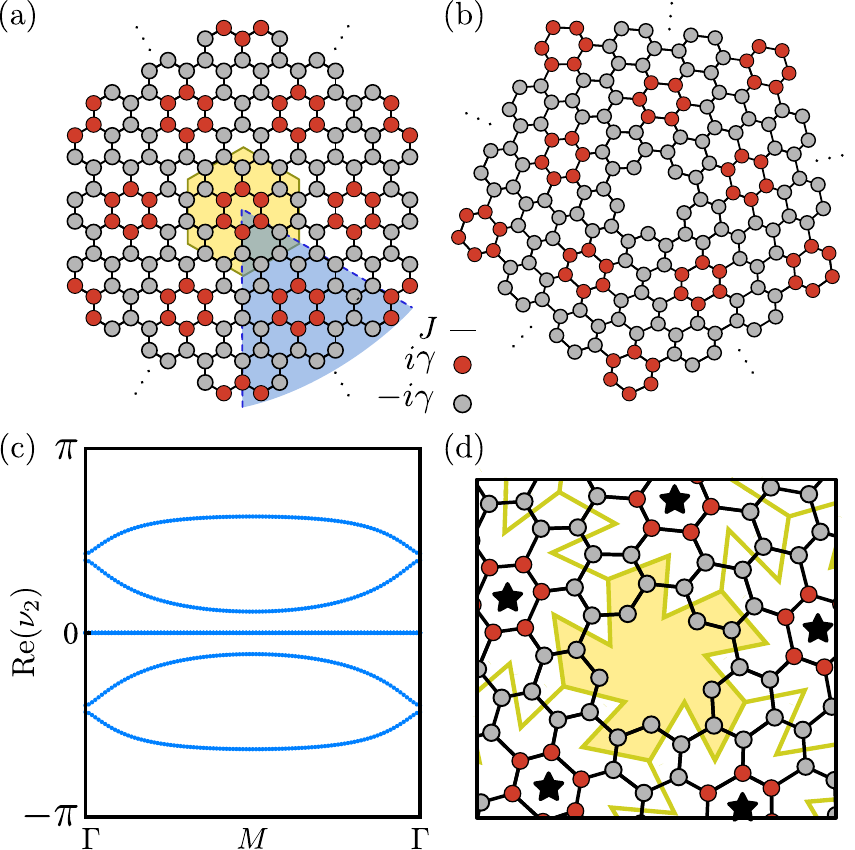}%
\caption{(a)--(b) An alternative NH lattice, similar to Figs.~\ref{Fig:Main1}(a) and \ref{Fig:Main2}(a), but with a different gain/loss pattern. (c) Wilson loop for the lowest nine bands, with $J=1$ and $\gamma=0.75$.  (d) Close-up view of the disclination-bearing lattice, with the Wannier centers marked by black stars.}
\label{Fig:Main3}
\end{figure}

\begin{figure}
\centering
\includegraphics[width=0.46\textwidth]{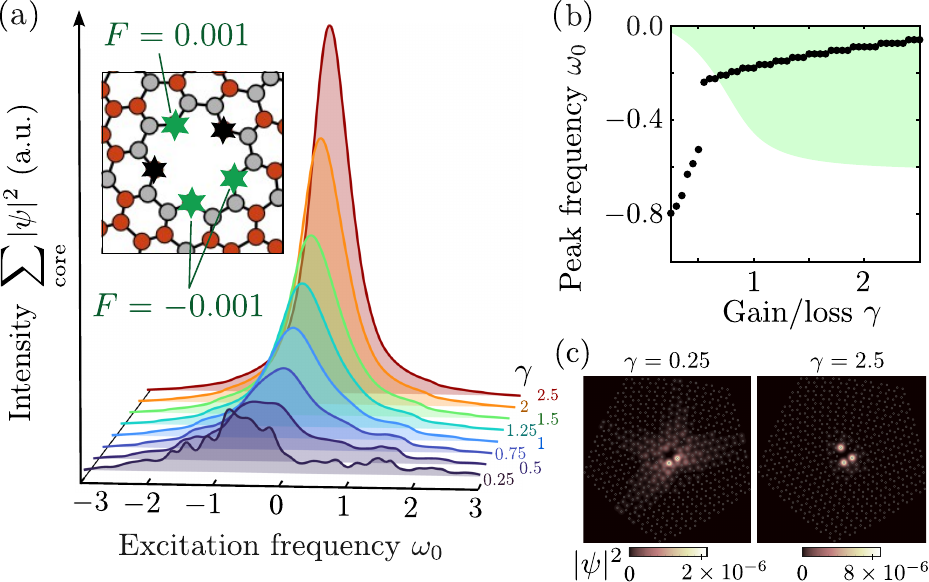}%
\caption{(a) Time domain simulation results for a disclination-bearing lattice [Fig.~\ref{Fig:Main1}(a)] driven by a monochromatic excitation on three sites at the disclination core (green stars, inset), with $F = 0$ on all other sites.  The total intensity on the five core sites (stars, inset) is plotted against the excitation frequency $\omega_0$ for different $\gamma$, with $J = 1$.  (b) Plot of peak frequency versus $\gamma$.  The shaded green area indicates the bulk gap of the undriven lattice.  (c) Steady-state intensity profile for two values of $\gamma$.  In the large-$\gamma$ regime, the intensity is strongly localized to the disclination core.}
\label{Fig:Fig4_Main}
\end{figure}


\textit{Alternative gain/loss pattern}---Fig.~\ref{Fig:Main3}(a)--(b) shows a lattice with an alternative gain/loss distribution.  Like the previous design, the periodic lattice has 18 sites per unit cell, and the gain/loss pattern produces no seam under the Volterra process. For $\gamma \ne 0$, this lattice exhibits a real line gap, but no corner or disclination states (for details, see the Supplemental Materials \cite{SuppMat}).  We thus interpret this as a NH analogue of a HOTI in the trivial phase.  Moreover, if we apply a disclination with a Frank angle of $2\pi/3$, the resulting $C_4$-symmetric lattice \textit{does} host a midgap disclination state \cite{SuppMat}, consistent with Deng \textit{et al.}'s finding that a Wu-Hu lattice in its trivial phase hosts a disclination state for a $2\pi/3$ disclination but not a $\pi/3$ disclination \cite{Deng2022Observation}.

It is also instructive to perform the Wilson loop calculation for this case. As shown in Fig.~\ref{Fig:Main3}(c), we obtain trivial windings reminiscent of the trivial phase of the Hermitian Wu-Hu model with Wannier centers at the $1a$ Wyckoff position \cite{Ahn2019Stiefel, Xu2023Absence}.  Fig.~\ref{Fig:Main3}(d) shows the positions of the Wannier centers (black stars), which lie at the centers of the unit cells and thus do not contribute fractional charge to the disclination core.  This is consistent with the lattice's aforementioned lack of gain/loss-induced disclination-bound states, and further supported by the calculation yielding $Q_d = -0.08$.

\textit{Experimental signatures}---The present model should be experimentally realizable using classical-wave platforms, such as arrays of photonic resonators \cite{Gao2020Observation, Liu2020Gain, Gao2021Non-Hermitian, liu2023localization, Ghatak2020Observation, Ota2020Active}.  The model's inter-site hoppings are reciprocal, positive, and nearest-neighbor, whereas the non-Hermiticity enters only as on-site gain and loss.  A possible complication is that inter-site hoppings in a real lattice may be spatially inhomogeneous due to varying physical distances between resonators.  However, our numerical studies show that such inhomogeneities do not substantially affect the disclination states \cite{SuppMat}.

As an example of how the NH disclination states could be identified in an experiment, we insert the lattice of Fig.~\ref{Fig:Main1}(a) into a driven Schr\"{o}dinger equation,
\begin{align}
i\frac{\partial \psi}{\partial t}=H\psi-i\gamma \psi+Fe^{-i\omega_0 t},
\label{eq:EqSch}
\end{align}
where $\psi$ is the state vector, $H$ is the NH lattice Hamiltonian, and an additional loss $\gamma$ is applied to all sites.  This can describe a setup where individual resonators have fixed loss and tunable gain, with the latter tuned so that one sublattice has zero net gain/loss and the other has loss $2\gamma$.  Eq.~\eqref{eq:EqSch} also includes an excitation of frequency $\omega_0$ and spatially-dependent amplitude $F$, concentrated on three disclination core sites as indicated in the inset of Fig.~\ref{Fig:Fig4_Main}(a) (overlapping one of the disclination states).

Fig.~\ref{Fig:Fig4_Main}(a) shows the resulting total intensity on the five disclination core sites, versus the excitation frequency $\omega_0$.  As we tune $\gamma$ to larger values, the intensity peak shifts closer to mid-gap and increases greatly in magnitude.  As shown in Fig.~\ref{Fig:Fig4_Main}(b), the peak frequency exhibits an abrupt jump, coinciding with the emergence of the disclination state into the theoretically-predicted gain/loss-induced bulk gap.  The steady-state intensity profiles, plotted in Fig.\ref{Fig:Fig4_Main}(c), show strong localization in the large-$\gamma$ regime.

{\textit{Conclusion}---}We have shown that disclinations in 2D lattices can host non-Hermitian disclination states, which emerge when a bulk gap is created by a specific pattern of gain/loss on the lattice sites.  The disclinations are shown to carry fractional charge $1/2$, thereby extending recent results on Hermitian topological disclination states into the non-Hermitian case.  Because these topological disclination states are induced by gain/loss tuning, they might be used as the basis for novel lasers whose modes can be manipulated via selective pumping.


This work was supported by the National Research Foundation (NRF), Singapore under Competitive Research Programme NRF-CRP23-2019-0005 and NRF-CRP23-2019-0007, and NRF Investigatorship NRF-NRFI08-2022-0001.

\bibliography{latex}

\end{document}